\documentclass[12pt]{iopart}

\usepackage[left,modulo]{lineno}
\usepackage{iopams}
\usepackage{graphicx}
\usepackage{color}
\usepackage{ulem}
\usepackage{multirow}
\usepackage{amssymb}
\usepackage{amsthm}
\usepackage{amsfonts}
\usepackage{textcomp}

\begin{document}

\title{Calibration of a superconducting gravimeter with an absolute atom gravimeter }

\author{S. Merlet$^1$\footnote{Orcid: 0000-0002-4746-2400}, P. Gillot$^1$,  B. Cheng$^1$\footnote{Present address: Institute of Optics, The Zhejiang Provincial Key Laboratory of Quantum Precision Measurement, College of Science, Zhejiang University of Technology, Hangzhou 310023, China}, R. Karcher$^1$, A. Imanaliev$^1$\footnote{Present address: Laboratoire National de M\'etrologie et Essais (LNE), 29 avenue Roger Hennequin, 78197 Trappes cedex, France}\footnote{Orcid: 0000-0002-8397-6927}, L. Timmen$^2$\footnote{Orcid: 0000-0003-2334-5282} and  F. Pereira Dos Santos$^1$\footnote{Orcid: 0000-0003-0659-5028}  }

\address{$^1$LNE-SYRTE, Observatoire de Paris - Universit\'e PSL, CNRS, Sorbonne Universit\'e,
61 avenue de l'Observatoire, 75014 Paris, France.
\newline
$^2$  Institute of Geodesy, Leibniz University Hannover, Scheinderberg 50, 30167 Hannover, Germany}

\ead{sebastien.merlet@obspm.fr}

\begin{abstract}
We present a 27-days long common view measurement of an absolute cold atom gravimeter (CAG) and a relative iGrav superconducting gravimeter, which we use to calibrate the iGrav scale factor. This allowed us to push the CAG long-term stability down to the level of 0.5~nm.s$^{-2}$. We investigate the impact of the duration of the measurement on the uncertainty in the determination of the correlation factor and show that it is limited to about 3\textperthousand~by the coloured noise of our cold atom gravimeter. A 3-days long measurement session with an additional FG5X absolute gravimeter allows us to directly compare the calibration results obtained with two different absolute meters. Based on our analysis, we expect that with an improvement of its long term stability, the CAG will allow to calibrate the iGrav scale factor to better than the per mille level (1$\sigma$ level of confidence) after only one-day of concurrent measurements for maximum tidal amplitudes.

%
\end{abstract}

\section{Introduction}
\label{intro}
Because of their high sensitivity, low drift and reasonable maintenance costs, superconducting gravimeters (SG)~\cite{Goodkind1999} are today the key instruments for the continuous monitoring of gravity variations. Nevertheless, being relative meters, they need to be calibrated and their drift to be determined, the methods for this being summarised for instance in Refs.~\cite{Imanishi2002} and \cite{Rosat2009}. For their calibration, one can either use long tidal measurements~\cite{Melchior1994}, induce controlled gravity changes by displacing masses, or the SG itself~\cite{Warburton1975}, \cite{Achilli1995}, \cite{Richter1995}, \cite{Falk2001}, perform co-located measurements with relative spring gravimeters~\cite{Meurers2002}, \cite{Meurers2012} or with absolute gravimeters (AG)~\cite{Hinderer1991}, \cite{Francis1997}, \cite{Francis1998}, this last method being today the most common. Although less precise in the end than using a calibration platform~\cite{Falk2001}, it has the advantage that it does not require moving the SG. Moreover, this is the only method which allows in addition to evaluate precisely the SG drift. With free fall corner cube AGs, such as the FG5s~\cite{Niebauer1995}, a precision of 1\textperthousand ~is typically reached in less than a week of concurrent measurement~\cite{Francis1998}, \cite{Imanishi2002}, \cite{Tamura2005}, \cite{Fuduka2005}, \cite{Rosat2009}, \cite{Hinderer2015}.

For applications in geophysics~\cite{Crossley1999}, \cite{Baker2003}, \cite{Meurers2016}, \cite{VanCamp2017} the accurate determination of the SG scale factor is important, and a long term stability of the gravity measurements is desirable. This motivates the regular intercomparison of SGs with AGs in order to track SGs drifts, potential changes in their scale factor, as well as offsets related to maintenance operations, or uncontrolled systematic effects.

Atom gravimeters based on atom interferometry~\cite{Kasevich1991} offer new measurement capabilities, by combining high sensitivities~\cite{Peters2001}, \cite{Hu2013}, \cite{Gillot2014}, \cite{Freier2016} and accuracies at the best level of a few tens of nm.s$^{-2}$~\cite{Peters2001}, \cite{Merlet2010}, \cite{Karcher2018}, with the possibility to perform continuous measurements~\cite{Peters2001}, \cite{LouchetChauvet2011}, \cite{Hauth2014}, \cite{Wang2018}, \cite{Menoret2018}. Being absolute meters, their scale factor is accurately determined and do not need calibration. This is essential for applications in the frame of fundamental metrology, such as for the determination of the Planck constant with a Kibble balance~\cite{Matthieu2017} and new realisation of the kilogram in the revised International System of Units~\cite{Stock2018}. As AGs, they are expected to play a role in the quest for the establishment of a global absolute gravity reference system and frame~\cite{Wziontek2021}. The study of long-term stability of atom gravimeters, as for any type of AGs, requires the precise knowledge of temporal fluctuations of gravity, in order to be able to separate them from fluctuations of systematic effects in the sensors. Even the best tide models are not enough for that purpose, as they do not account for all processes that do change the local value of gravity. This prevented us for a long time to assess the long-term stability of our Cold Atom Gravimeter (CAG), such as in Ref.~\cite{LouchetChauvet2011} where one could not assess whether the long-term stability of a 12-days continuous gravity measurement was limited by the instrument or by the tidal model. In 2013, the comparison of our measurements with an improved tidal model allowed to demonstrate a stability of 2~nm.s$^{-2}$ at 1~000~s measurement time~\cite{Gillot2014}. Nevertheless, direct comparisons between different sensors, and in particular with SGs, is preferable. In 2015, the GAIN gravimeter of Humboldt-Universit\"at zu Berlin reached an unprecedented stability of 0.5~nm.s$^{-2}$ in 10$^5$~s of measurement time when compared to an Observatory SG (OSG)~\cite{Freier2016}.

Since the very beginning of 2013, we operate a superconducting gravimeter in our gravity laboratory at LNE~\cite{Merlet2008}. The CAG also operates in this laboratory since the end of the CIPM Key Comparison CCM.G-K1 which occurred during ICAG'09~\cite{Jiang2012}. Since then, we have been continuously working at improving the instrument performance, except when participating to comparisons in other laboratories~\cite{Francis2013}, \cite{Francis2015} or demonstrating the capabilities of atom interferometers~\cite{Farah2014} in the LSBB underground facility for the MIGA project~\cite{Geiger2015}. We performed a one-month long gravity measurement performed simultaneously by the iGrav005\footnote{\label{note} doi.org/10.5880/igets.tr.l1.001} and the CAG, to investigate the long term stability of our atomic instrument. We present in this paper this measurement and take advantage to study the calibration of the SG and discuss the uncertainty of this calibration.

\section{Continuous common view gravity measurement with atom and superconducting gravimeters}
\label{sec:1}
\begin{figure}
\includegraphics[width=0.7\textwidth]{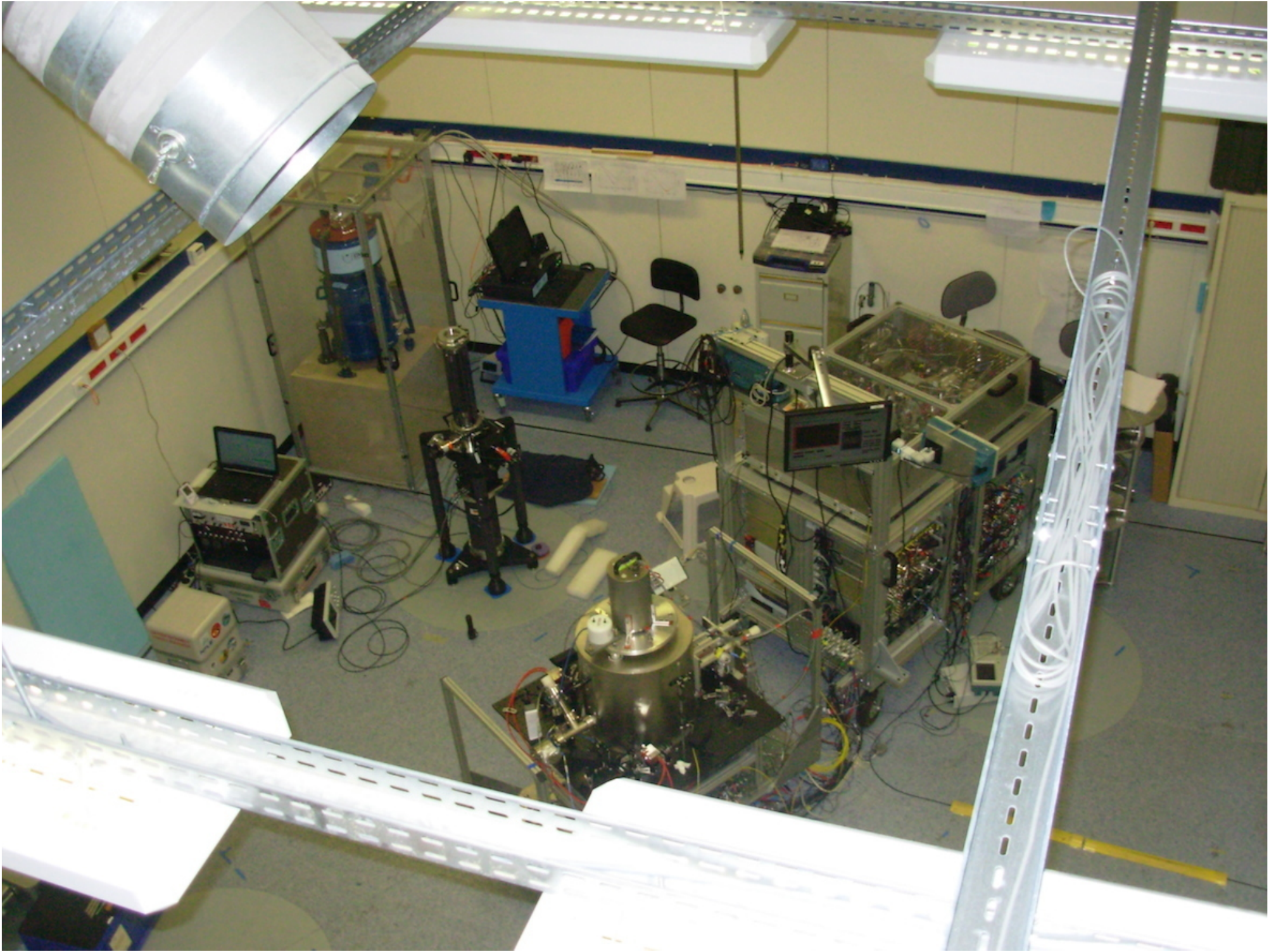}
\caption{Picture of the LNE gravimetry laboratory. A 6~m~$\times$~5.5~m pillar can host several gravimeters at a time for calibration and comparison campaigns. The iGrav005 is installed onto a cubic stone attached to the laboratory pillar at one of its corners. The CAG is placed at the center of the pillar. The FG5X-220 of Institute of Geodesy of Leibniz University Hannover is installed at one of the four remaining measurement stations. }
\label{Labo}
\end{figure}

The LNE gravimetry laboratory is equipped with a 33~m$^2$ pillar, large enough to accommodate several AGs at a time for intercomparisons. A superconducting gravimeter (iGrav005)$^{\ref{note}}$ is operating since 2013 at one of the pillar's corners. It is placed on top of a rock pedestal, so as to raise the measurement height of the instrument at mid human height ($\sim$ 0.95~m) which reduces the gravity effect of the CAG operators. Figure~\ref{Labo} shows the two gravimeters, the iGrav005 on the stone and the CAG at the centre of the laboratory. They are located 3.5~m apart in the horizontal plane and their measurement height differs by 0.1~m. The direct comparison between these two instruments is of interest as they rely on very different measurement methods, and do not in principle suffer from common systematic effects that might exist when comparing instruments of the same technology, if not from the same family.

The iGrav output signal is the feedback signal that controls the levitation of a superconducting sphere in a magnetic field generated with superconducting coils~\cite{Goodkind1999}. When gravity changes, this output feedback signal, a voltage, is modified. As for the CAG, the signal of interest is a frequency chirp applied to the interferometer lasers, which continuously stirs the phase of the interferometer to the center of the fringe pattern ~\cite{LouchetChauvet2011}, \cite{Farah2014}. The CAG has already been presented in detail, for instance in~\cite{LouchetChauvet2011}. We recall here the features mostly relevant for this study. Its measurement height is 0.835~m. In short, it performs cyclic measurements of $g$, with a cycle time of 380~ms. Its principle of operation is based on the realisation of a Mach Zehnder type atom interferometer with $^{87}Rb$ cold atoms. A measurement protocol, also described in~\cite{LouchetChauvet2011}, and based on alternating 4 different measurement configurations, allows to reject most of the systematics. Each measurement configuration is repeated 100 times after which a number of relevant parameters, such as tilts, optical powers and frequency quartz reference for example, are monitored. It leads to a gravity measurement average over 177~s. The CAG accuracy has recently been improved to 20~nm.s$^{-2}$~\cite{Karcher2018} but was in the range 40-50~nm.s$^{-2}$ for the measurements presented in the following.

\begin{figure*}[h!]
\centering
\noindent\includegraphics[width=1\textwidth]{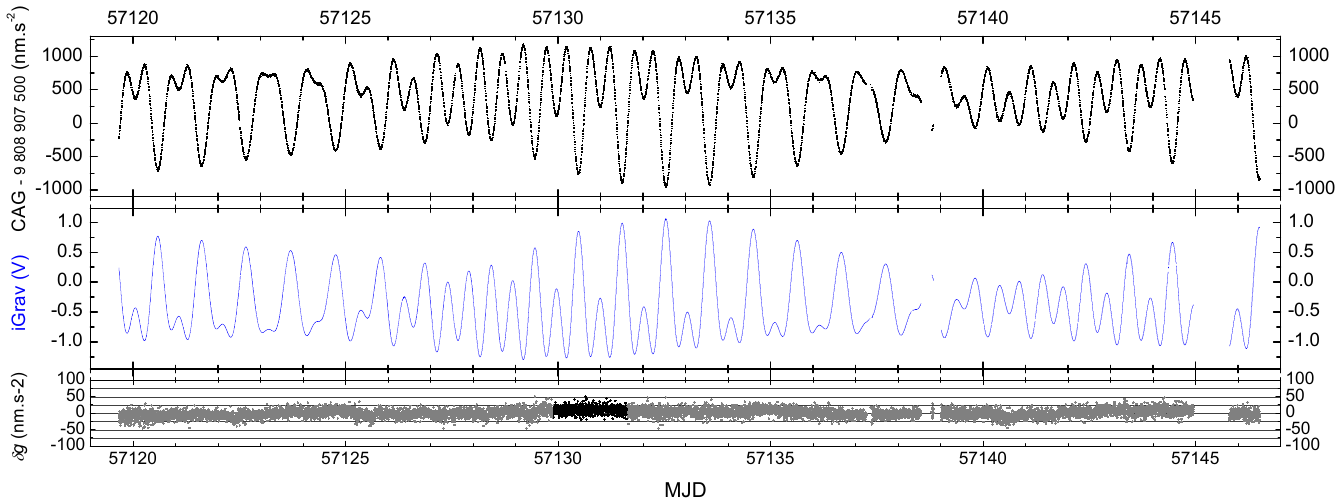}
\caption{Continuous gravity signals as measured by the CAG (in black) and the iGrav005 (in blue) from April the $7^{th}$ to May the $4^{th}$ of 2015. Data are averaged over the same duration of 177~s for both instruments. The difference $\delta g$ between the two instrument measurements after the calibration of the iGrav scale factor is represented in grey on the bottom graph ($\delta g = g_\mathrm{CAG}-g_\mathrm{iGrav}$). A shorter sample spanning over a week-end is highlighted in black. }
\label{measures}
\end{figure*}

Figure~\ref{measures} displays the results of a continuous gravity measurement session performed with the two instruments (CAG and iGrav) from April the $7^{th}$ to May the $4^{th}$ of 2015. Signals severely perturbed by the occurence of an earthquake, or by a failure of the CAG due to lasers going out of lock have been removed, leading to a few gaps in the data. The iGrav005 data are corrected for a linear drift of (0.027~$\pm$~0.006)~nm.s$^{-2}$.d$^{-1}$ evaluated over 4 years from 2014 to 2018.

\section{Instrumental delays}

As for tidal analysis~\cite{Baker2003}, an accurate timing of the data is required when comparing the gravity variations of the two instruments to calibrate the iGrav output signal~\cite{Meurers2002}, \cite{VanCamp2016}. The effect of a lack of synchronisation on the calibration of the SG depends on the amplitude of gravity variations and the duration of the common view measurement. For a 11 days session, which allows to observe tides with large and small amplitudes, with 30~s averaged data to filter and decrease the noise, a 30~s difference between the two instrument timings leads to an effect lower than 0.1\textperthousand~on the calibration factor (CF) determination such as results of~\cite{Meurers2002}, \cite{VanCamp2016}. For a one day session, the effect varies from 0.2\textperthousand~to 0.01\textperthousand~depending on the magnitude of the tides. The time stamping of CAG data is performed by the clock of its control computer, which is locked on UTC via the NTP protocol. The iGrav SG uses a GPS receiver to correct GPS time into UTC. In addition, one should also take into account delays due to the time response of the sensors to gravity changes. While the CAG suffers negligible delay when considering the time of the measurement at the middle of the interferometer, appreciable delays are present in the case of the SG, owing not only to their mechanical response function but also on the use of additional filters. A precise determination of the response function can in principle be performed via self calibration, but this functionality was not available in our iGrav. While the theoretical transfer function given in the iGrav Operator Manual allows us to estimate a delay of 10.9~s, Ref.~\cite{Francis2011} points out the need for considering instead real transfer functions, which can differ for each SG. A rough confirmation of this delay was obtained via the direct measurement of the gravity change when bringing suddenly 4 operators next to the meter and having them sit on the floor. We observed a gravity change by (11.6~$\pm$~0.3)~nm.s$^{-2}$ with a delay of (11~$\pm$~1)~s, in agreement with the theoretical estimation. Finally, this delay could also be extracted from the analysis of the time correlation between the signals of the two meters of figure~\ref{measures}, which allows for the determination of a delay of (10.3~$\pm$~0.3)~s. Note that these last two methods determine the overall SG delay, including additional delays by the data acquisition system \cite{VanCamp2000}.

\section{Direct calibration of the superconducting gravimeter}

Methods to calibrate SGs have already been investigated in detail in~\cite{Hinderer1991} \cite{Francis1998} and later improved in~\cite{VanCamp2016}. The relationship between the two signals (expressed in nm.s$^{-2}$ for the CAG and in volts for the iGrav) can simply be determined via a simple linear regression:

\begin{equation}\label{eq1}
\mathrm{AG} = \mathrm{CF}~\times~\mathrm{SG}~+~\mathrm{Offset}
\end{equation}

In ref.~\cite{Hinderer1991} a linear drift was added to take into account possible drift between AG and SG observations. When using new generation AGs (model FG5) refs.~\cite{Almavict1998} and \cite{Francis1998} did not include any additional linear or polynomial part, but based the CF determination on equation~\ref{eq1}. When the calibration involve such as a calibrated SG versus a spring gravimeter to be calibrated for instance, the use of polynomials is required~\cite{Francis2001}. The polynomial term has the advantage of removing any drift of the SG or any residual drift after an imperfect drift removal. It can then also in principle remove drifts due to fluctuations of uncontrolled systematic effect of the AG. We will discuss them in section~\ref{SecInstrumFluc}. In such a case, the polynomial term prevent to discriminate the relative contribution of the two instruments in the adjusted drift. Given that we are interested in assessing the potential of high stability atom gravimeters for the purpose of calibrating SGs, we choose first not to use additional polynomials for our study, though it is a common practice, which was shown to improve the CF determination in Ref.~\cite{Meurers2002}.

\begin{figure}[h!]
\includegraphics[width=0.7\textwidth]{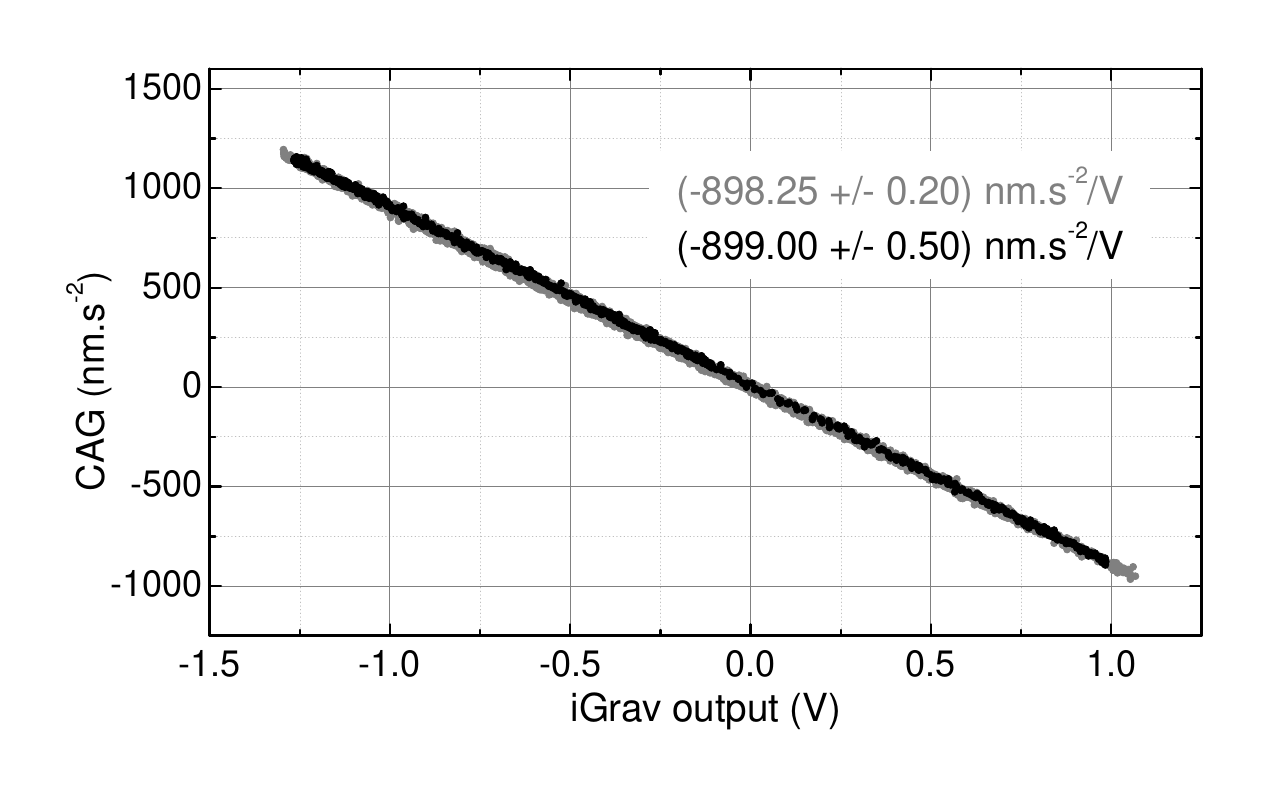}
\caption{Calibration of the iGrav output signal for the whole 27-days long measurement period (in grey) and for a selected shorter 1.7-day long period (in black). }
\label{Calib}
\end{figure}

\begin{figure}[h!]
\includegraphics[width=0.7\textwidth]{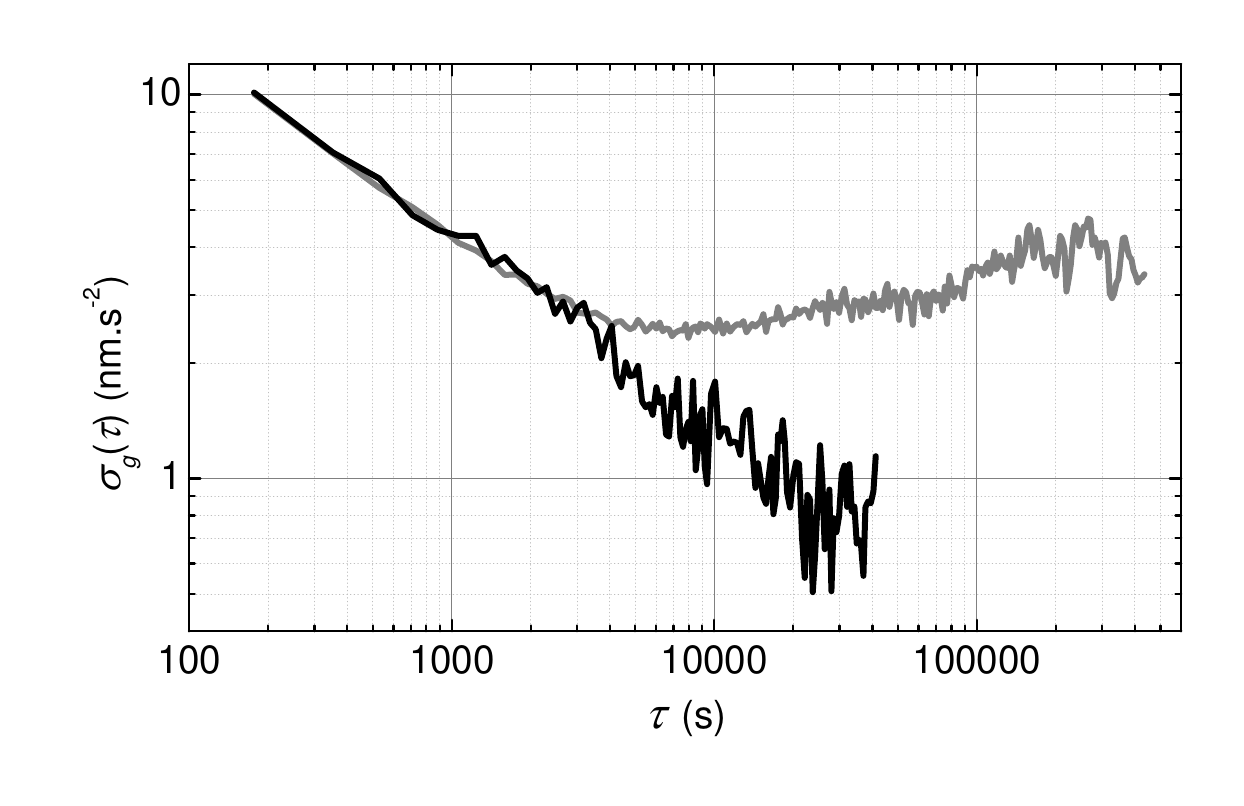}
\caption{Allan Standard deviations of the difference between the CAG and iGrav gravity signals, for the whole 27-days long measurement period (in grey) and for a selected shorter 1.7-day long period (in black).}
\label{figvar}
\end{figure}

Figure~\ref{Calib} illustrates the correlation between the measurements of the two instruments. We used equation~\ref{eq1} to determine the CFs over two sets of data of different lengths. The first set displayed in grey corresponds to the whole measurement period of 27-days, while the second in black corresponds to a more quiet period of 1.7-days, starting before midnight on a Friday and ending after midday on the next Sunday, the noise during weekends being reduced by the absence of on site human activity. We obtain two calibration factors of respectively (-898.25~$\pm$~0.20)~nm.s$^{-2}$/V and (-899.00~$\pm$~0.50)~nm.s$^{-2}$/V, in agreement within their uncertainties, which are given here by the errors of the fits (1$\sigma$), ie the standard errors of the regression slopes. The first calibration factor was then used to convert the SG voltage samples into gravity data, and the difference between CAG and the calibrated SG measurements was calculated. This difference is displayed in the bottom part of figure~\ref{measures}. A statistical analysis of this difference over the two sets of data was then performed by calculating their Allan standard deviations, which are displayed on figure~\ref{figvar}. For the selected 1.7-day period, the Allan standard deviation averages down to 0.5~-~0.6~nm.s$^{-2}$, with a $\tau^{-1/2}$ slope characteristic of white noise, as already observed in \cite{Freier2016}. As for the Allan standard deviation of the 27-days common view measurement, it also decreases with the same slope down to the 2~-~3~nm.s$^{-2}$ level for about 3~000~-~4~000~s, but reaches a plateau for larger averaging times. 

\section{Segmented duration analysis }\label{sec segmentation}\label{sectseg}

\begin{figure}[h!]
\includegraphics[width=0.7\textwidth]{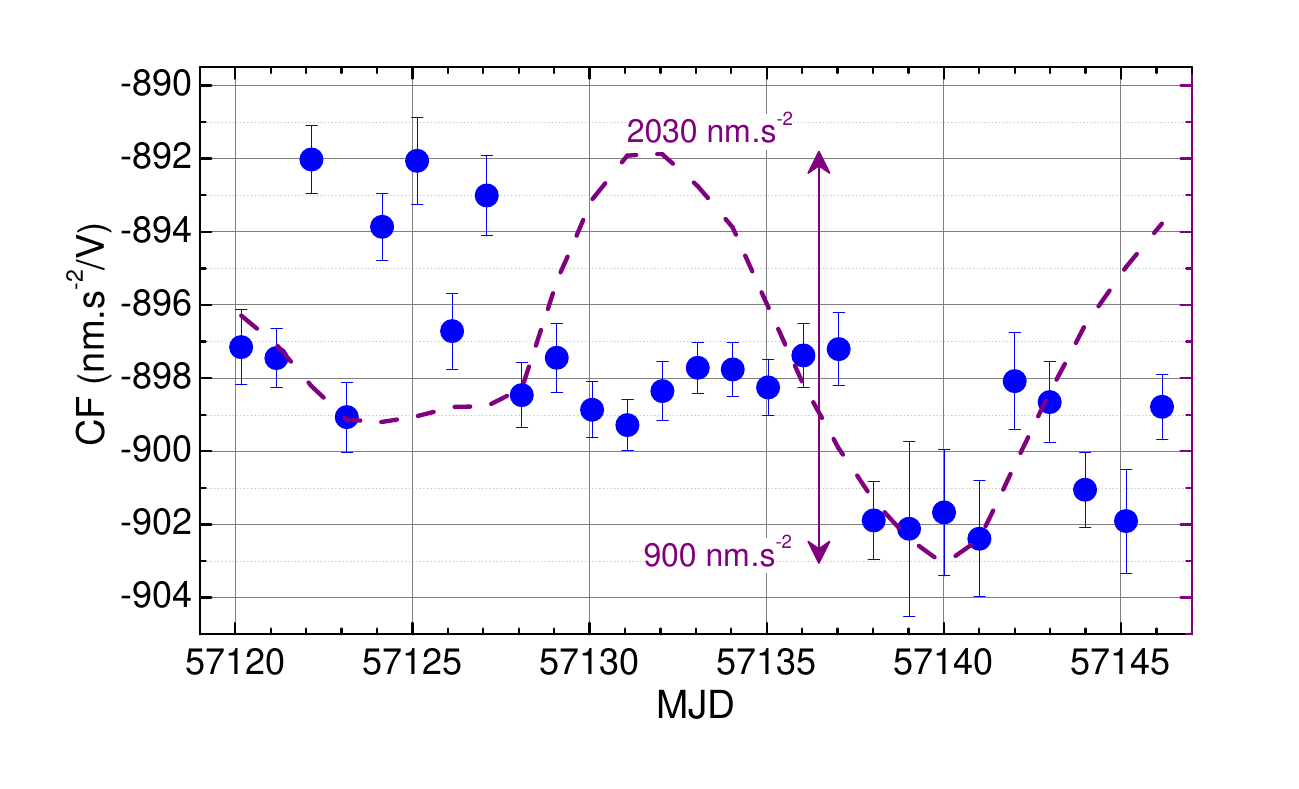}
\caption{iGrav one-day calibration factors. The error bars are the errors of the individual fits. The dashed purple line represents the peak-to-peak tidal gravity change over one day.}
\label{FigCFCAG1d}
\end{figure}

To perform a statistically relevant study of the uncertainty associated with the calibration factor determination, several independent such determinations would be required. We thus take advantage of the time-length of the measurement to carry out a segmented analysis and calculate an iGrav CF for each day of measurement. The 27 resulting one-day calibration factors are displayed in figure~\ref{FigCFCAG1d}, with uncertainties given by the errors of the linear fits. Remarkably, the standard deviation of the one-day CFs, which amounts to 2.77~nm.s$^{-2}$/V, is three times larger than the mean value of the errors of the fits of 1.05~nm.s$^{-2}$/V. This tends to indicate that the errors of the fits underestimate the uncertainty in the CF determinations. The peak-to-peak variation of the one-day CFs is 10~nm.s$^{-2}$/V, twice smaller than in Ref.~\cite{Imanishi2002}, where a similar analysis was carried out between a SG and a FG5 AG over a similar 27-days long period, with peak to peak gravity variations up to 2~400~nm.s$^{-2}$.

We stress here that the observed variations of the CFs are not correlated with changes of the signal to noise ratio (S/N) which we define here as the ratio between $i)$ the difference between maximal and minimal tidal values over the one-day long observation and $ii)$ the Allan standard deviation of the signal for an averaging time of 177~s, which we take here as the short term noise. Such variation can in practice impact the CF as discussed in Ref.~\cite{VanCamp2016}. Nevertheless, given that SG data are averaged over a relatively long duration of 177~s, the SG noise is here too small to lead to significant attenuations of the CFs, comparable to the amplitude of the fluctuation we observe on figure~\ref{FigCFCAG1d}. Yet, one can notice that the CF values determined for large peak-to-peak tide amplitudes, and in particular between day 57~129 and 57~135, show less scattering.

Here, as for the global CF determination of previous section, we used equation~\ref{eq1} for each day segment analysis. As discussed later, we repeated the protocole with a polynomial part added to equation~\ref{eq1} which did not modified the results whereas we used first or second order polynomial. 

\begin{figure}[h!]
\includegraphics[width=0.7\textwidth]{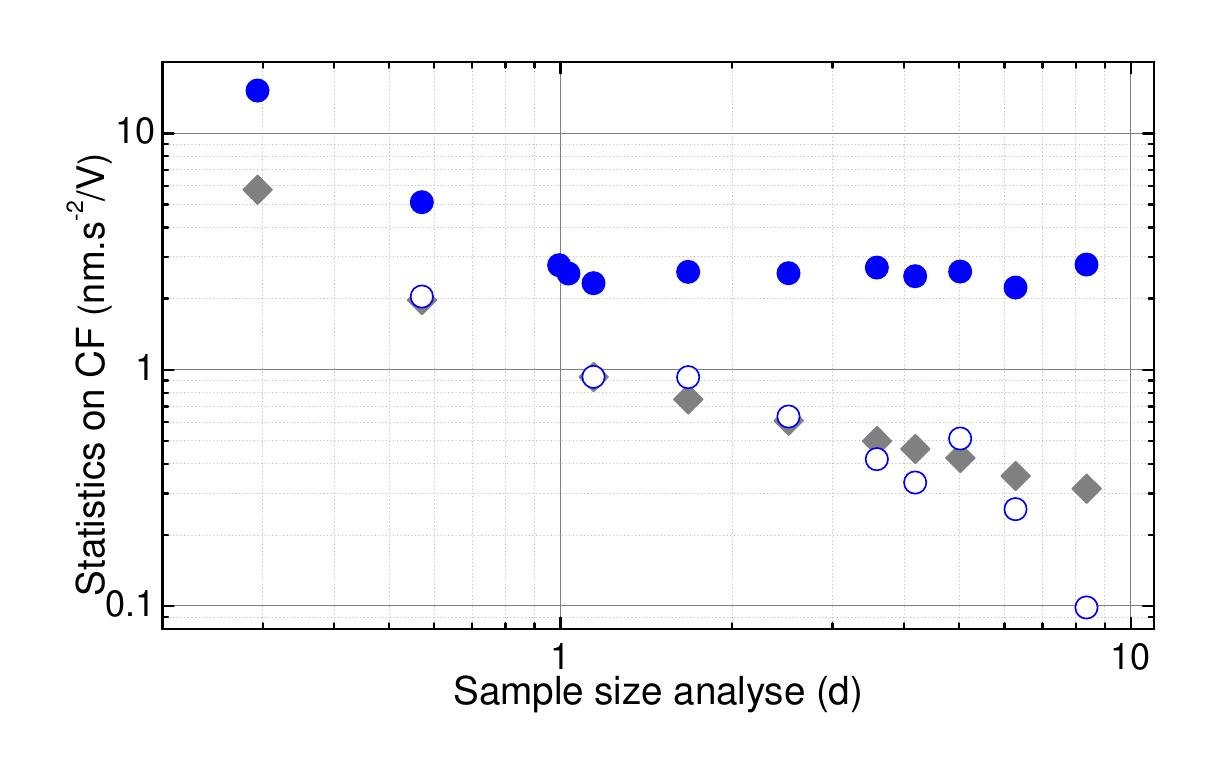}
\caption{Statistic analysis of iGrav CF determinations for different durations of measurement segmentation. Standard deviations of the distribution of the CFs are represented in blue dots: full dots for the real CAG data, and open dots for the synthetic AG signal. Grey diamonds display the means of the errors of the CF fits for the synthetic signal. }
\label{figsimus}
\end{figure}

To better understand the figure~\ref{FigCFCAG1d} behavior and compare the results we obtain with simulated data, we generated a synthetic AG signal, obtained out of the iGrav output signal converted into a gravity signal with the first CF obtained in figure~\ref{Calib}, to which a white noise of the same amplitude as the short term noise of the CAG was added. We then used this synthetic signal to calibrate the iGrav, repeating our segmented analysis, but for different measurement durations, spanning from 7~h to 200~h. As the total common view measurement is 27-days long, we obtained several distributions with numbers of samples ranging from 91 to 3 respectively. We report on figure~\ref{figsimus} the standard deviations of these distributions as open blue dots, as well as the corresponding means of the errors of the fits as grey diamonds, and we find a fair agreement between them. This shows that the mean errors of the fits are good estimates of the uncertainties in the CF determinations when the differential noise between the sensors is white. By contrast, the same analysis performed with the real CAG signal shows a different behavior. Indeed, the standard deviations, which are displayed as blue dots on figure~\ref{figsimus}, clearly feature a plateau, showing that measurements longer than about a day do not help to reduce the uncertainty on the CF determination. On the other hand, for durations of up to a day or so, the errors of the CF fits could be taken as reasonable estimators of the uncertainty of the CFs, despite being about twice overoptimistic. With this analysis, we understand that the behaviors we observe with the real data are related to coloured noise. Though at this stage, the question remains in principle open whether the coloured noise arises from the CAG or from the iGrav, we attribute it to the CAG. Indeed, the Allan standard deviations of the residuals of the gravity data corrected from tides, with a tidal model obtained with a spring gravimeter two years long continuous measurement session which improved the initial determination of Ref.~\cite{Merlet2008}, and from atmospheric effects, show for the CAG a behavior similar to the grey curve of figure~\ref{figvar} whereas, for the iGrav005, the stability is about three times lower at a 10~000~s averaging time.

\section{Comparison of calibrations with different types absolute gravimeters} \label{seccompAG}

In Ref.~\cite{Francis2002}, the authors calibrated a SG with four FG5 AGs during a single measurement session. The different CFs they obtained agreed with each other, demonstrating the robustness of using any FG5 AG for such calibration. Yet, one could not exclude a possible unaccounted-for bias because the experiment was limited to only FG5 AGs, which motivates carrying similar studies with AGs relying on different technologies. 

To do so, we took the opportunity of a measurement campaign organised in the frame of the ITOC project~\cite{Denker2018} to welcome again in the LNE gravimetry laboratory the free fall corner cube gravimeter FG5-220 of the Institute of Geodesy of Leibniz University Hannover, in its improved version~\cite{Niebauer2011}, namely the FG5X-220. As a remark, absolute gravity measurements performed by the AGs agree with a difference of 1~nm.s$^{-2}$ compatible with their measurement uncertainties. But, of particular interest for the present study, we took advantage of a week-end to perform a common view continuous measurement between the three instruments (iGrav, CAG, FG5X) to repeat the analysis presented above in section~\ref{sec segmentation}. Note that for these measurements, $i)$ the FG5X-220 performed one free fall every 30~s continuously during close to 3 days, a duration not far from the 4-5 days required to calibrate an SG at the 1\textperthousand~level with an FG5, according to \cite{Francis1998};  $ii)$ the CAG cycling $g$ measurement was 360~ms which led to $g$ measurement averaged over 164~s in the protocole summarised in section~\ref{sec:1} and $iii)$ the maximum gravity difference due to peak-to-peak tide variations was 1~500~nm.s$^{-2}$.

\subsection{Global and segmentation analysis}

\begin{figure}[h!]
\includegraphics[width=0.7\textwidth]{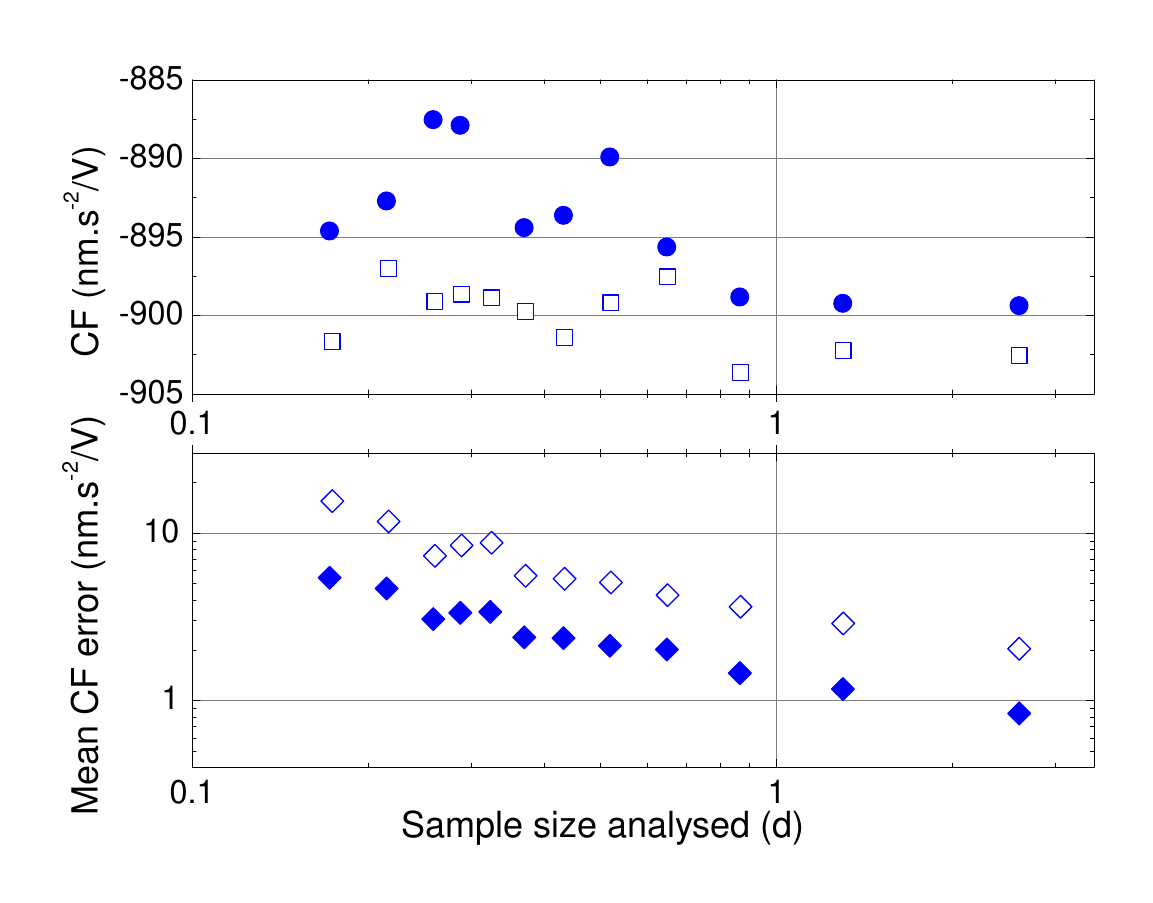}
\caption{Mean calibration factors of the iGrav005, and mean CF fit errors, obtained with the CAG (full dots and diamonds) and the FG5X-220 (opened squares and diamonds) for different durations of segmentation of the measurement.}
\label{Fig-CF-CAG-FG5}
\end{figure}

Figure~\ref{Fig-CF-CAG-FG5} presents the results of the statistical analysis of the iGrav CFs determined for segment sizes varying from 4 h to 62.5 h, corresponding respectively to numbers of independent determinations of the CF ranging from 15 to 1. We used the FG5X-220 drops and verified that the FG5X values were not affected by aliasing effect due to the 30~s measurement period~\cite{VanCamp2005}. The analysis leads to a CF difference of (3.2~$\pm$~2.2)~nm.s$^{-2}$/V for the whole measurement duration. The mean of the errors of the CF fit, which we take here as fair estimates of the uncertainties in the CF determinations, is three times better for the CAG determination due to its better short term sensitivity~\cite{Gillot2014}. Nevertheless, in principle these uncertainties associated to the FG5X could be reduced by increasing the repetition rate. Expressed as in many papers~\cite{Francis1998}, \cite{Imanishi2002}, \cite{Tamura2005}, \cite{Fuduka2005}, \cite{Rosat2009}, \cite{Meurers2012}, \cite{Crossley2018} in relative units, it corresponds to a relative uncertainty  of 0.7\textperthousand~for the CAG and 2.3\textperthousand~for the FG5X. The latter value being compatible with similar previous studies~\cite{Francis1998}.

\subsection{Focus on one-day segmentation}

\begin{figure}[h!]
\includegraphics[width=0.7\textwidth]{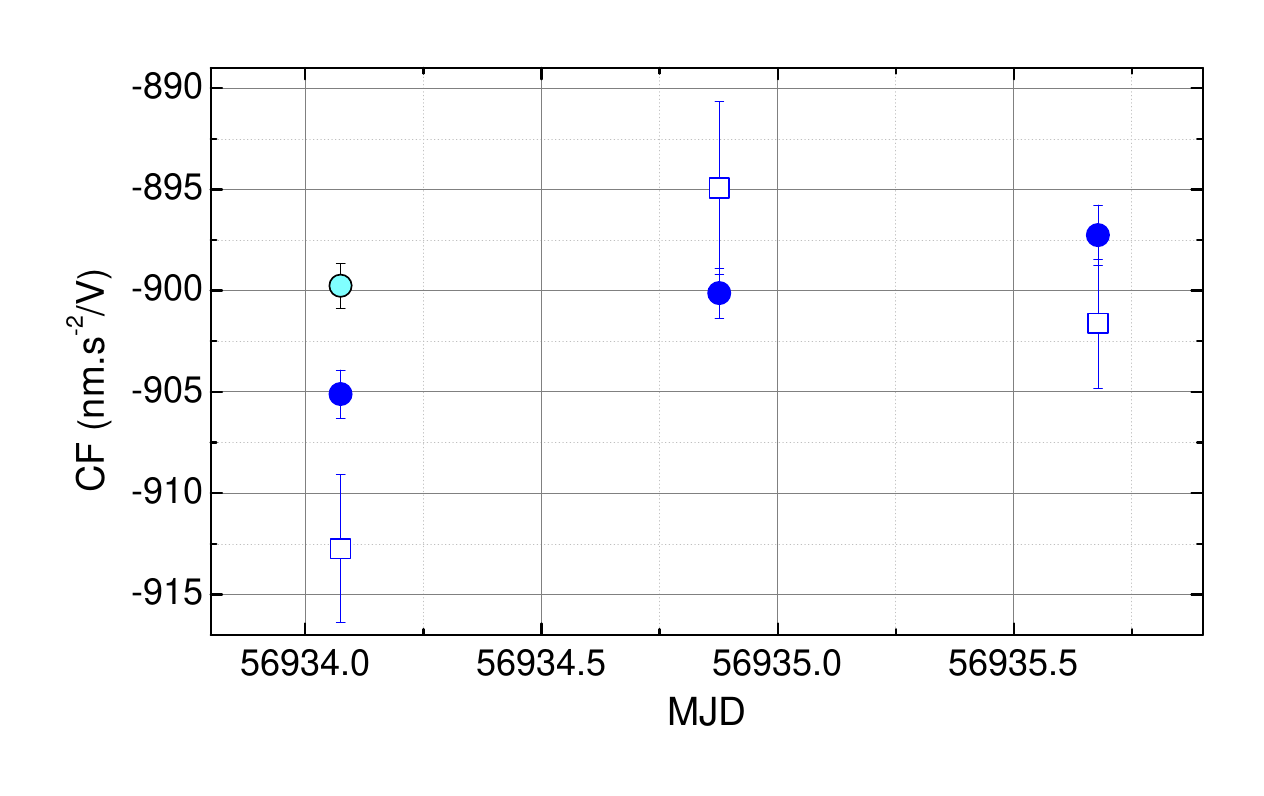}
\caption{iGrav calibration factors determined during three consecutive one-day common view measurements obtained with the CAG (blue dots) and FG5X-220 (opened squares). The 1$^{st}$ day result obtained with CAG measurement corrected from a linear drift (see Fig.~\ref{FigResiduals}) is represented with a cyan dot. The error bars are the individual fit errors. }
\label{Fig-CF-CAG-FG5-1d}
\end{figure}

As the measurement with the FG5X was not exactly 3-days long, we then split the measurement data into 3 slightly overlapping periods of 1 day length in order to perform three 1-day analysis such as the one presented in figure~\ref{FigCFCAG1d}. The overlap between two consecutive segments is of order of 10\%. Figure~\ref{Fig-CF-CAG-FG5-1d} displays the results of these 1-day analysis for both instruments. As in the analysis of section~\ref{sec segmentation}, the CFs vary from one day to the other: 7.5~nm.s$^{-2}$/V peak-to-peak with the CAG and 17.8~nm.s$^{-2}$/V with FG5X. As expected from the above global analysis we do not find statistically resolved differences between the CFs of the CAG and the FG5X, their differences being lower than their combined 2$\sigma$ uncertainties. Nevertheless, the two determinations of the first day lie a bit off with respect to the 2 others.

\section{Discussion of the results}

\begin{figure}[h!]
\includegraphics[width=0.7\textwidth]{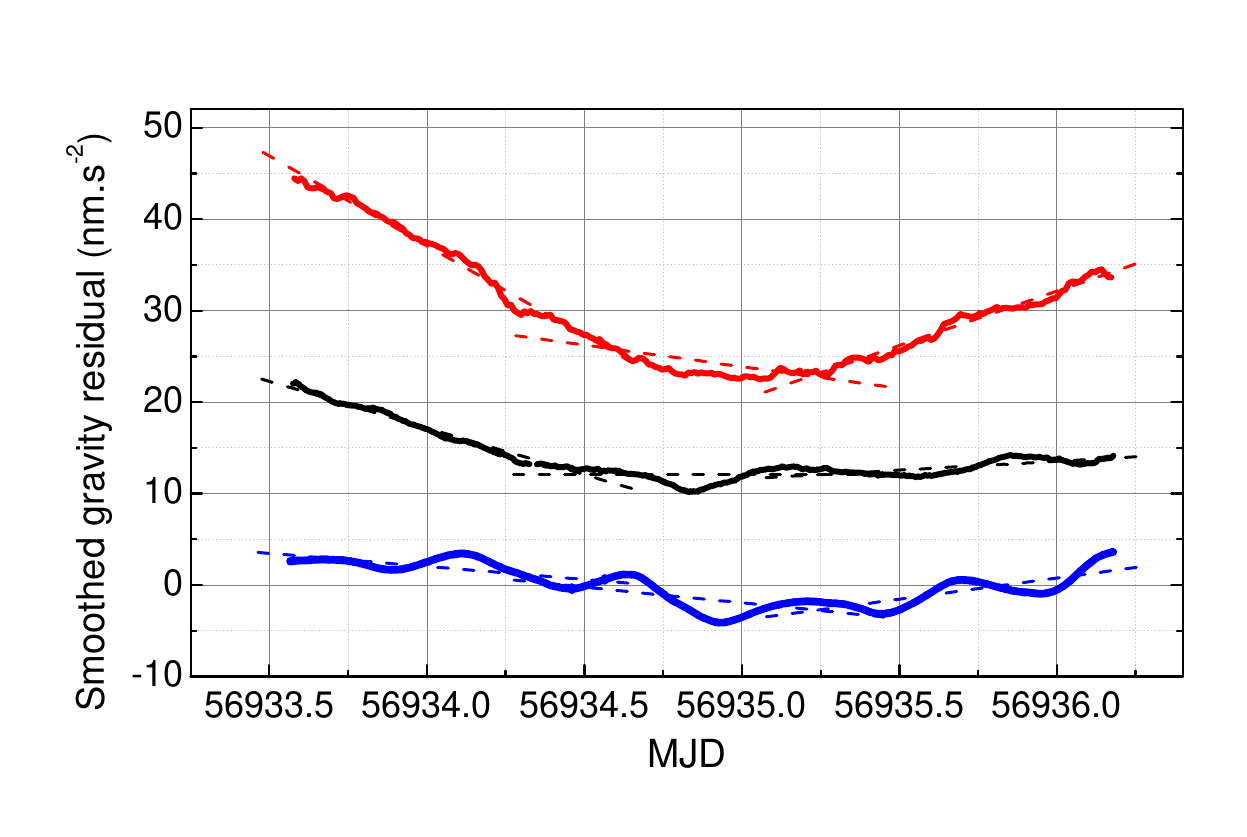}
\caption{Gravity variations corrected for tides, atmospheric pressure and polar motion effects during the 2014 common view measurement. iGrav005 residual is represented in blue, FG5X-220 is red and CAG in black. The three individual smoothed residuals are shifted for legibility. The dashed lines represent 9 individual linear drifts calculated for each one day measurement of each instrument. }
\label{FigResiduals}
\end{figure}

In principle, differences in the CFs represented in figure~\ref{Fig-CF-CAG-FG5-1d} could be explained either by $i)$ real gravity fluctuations, which are not perfectly correlated between the instruments, $ii)$ instability of the SG, $iii)$ instability of the AGs. The proximity between the sensors (1.4~m between iGrav and FG5X-220, 2.1~m between FG5X-220 and CAG and 3.5~m between iGrav and CAG) rules out the first hypothesis. For the second, even if SGs have been shown to be stable \cite{Rosat2018}, we cannot at this stage exclude a problem with ours. To investigate this, we calculated CFs by comparing SG data corrected from pressure and polar motion effects with the tidal model improved from~\cite{Merlet2008} presented in section~\ref{sectseg}. We then found the three one-day determinations in agreement within 0.8~nm.s$^{-2}$/V, with an average value of (-897.3~$\pm$~0.3)~nm.s$^{-2}$/V, confirming the good behavior of the SG. Finally, gravity measurements from the three meters were corrected from the same tidal model and other corrections mentioned above, leading to the residuals displayed in figure~\ref{FigResiduals}. For a better readability, the three curves are offset by 15~nm.s$^{-2}$. While SG residuals can be attributed to imperfect tidal predictions and corrections, the AG residuals display clear drifts which we attribute to bias instabilities. Residuals for both AGs do show a similar trend (a linear drift) at the beginning, while after about a day, the CAG residual remain more or less stable while the FG5X residual drifts in the opposite direction. We calculated daily linear drifts for each instruments, which are displayed as dashed lines in figure~\ref{FigResiduals}. For instance a drift of (-10.19~$\pm$~0.06)~nm.s$^{-2}$.d$^{-1}$ was found  over the first day for the CAG. Correcting CAG data from this drift changes the first CF value by 5.4~nm.s$^{-2}$/V, and brings it as expected closer to the two others (see the point in cyan in figure~\ref{Fig-CF-CAG-FG5-1d}). A similar analysis with FG5X show the same behavior. When correcting from a quadratic drift the global 3 days analysis, we reduce the difference between the CFs down to (1.2~$\pm$~2.2)~nm.s$^{-2}$/V. This clearly shows that the determinations are affected by the instabilities of the AGs.

When the fluctuation is "deterministic", such as in the case of a linear (or eventually polynomial) trend, one can account for it in the fit formula for the CF determination~\cite{Hinderer1991}. While this turns efficient in the example above, we found no significant difference in the dispersion of the figure~\ref{measures}  CFs for the 27-days measurement when adding in the fit a drift, be it linear or quadratic. This arrises from the fact that AGs instabilities are not necessarily well captured by linear or polynomial trends. Nevertheless, for peak-to-peak gravity variations ranging from 900~nm.s$^{-2}$ to 2~000~nm.s$^{-2}$ over a day, we find with the CAG that error fit represented in figure~\ref{figsimus} could lead to CF relative uncertainty better than the 1\textperthousand~level (1$\sigma$) as early as one day of measurement. This is all the more true for higher gravity variations.

More generally, our analyse is well in line with the recent study of Ref.~\cite{Scherneck2020} which highlights the crucial impact of long-term instabilities of AGs for SG calibrations and also shows the absolute atom instruments capabilities and the potential they offer for the determination of precise SG CFs. In particular the unprecedented measurement stability of 0.5~nm.s$^{-2}$ obtained with an AG~\cite{Freier2016} allows for determining the CF with a relative uncertainty of 0.3\textperthousand~in a 5.8~d measurement during peak-to-peak gravity variations of 1~200~nm.s$^{-2}$.

\section{Instrumental fluctuations}\label{SecInstrumFluc}

We finally investigate the sources of the drifts observed in the residuals represented in figure~\ref{FigResiduals}, by looking at correlations with relevant experimental parameters.

We found no correlation between the residuals and any of the parameters monitored for the FG5X. However this instrument class has a tendency to lose its verticality due to temperature variation for exemple, which thus decrease gravity value. This parameter is not monitored during measurements. Despite the laboratory temperature varied of 0.15$^{\circ}$C peak-to-peak during the measurement, we cannot exclude a thermal and/or mechanical effect relaxation on the alignments due to operator manual adjustments on the sensor. The verticality could have varied during the week-end measurement and came back to the initial value when we enter back in the laboratory.

The trend in the CAG data is well correlated with a drift in the optical powers of the lasers used to drive the interferometer. The evolution of these laser optical powers leads to an evolution of the so called two photon light shift (TPLS)~\cite{Gauguet2008}. It drifted linearly with a trend expressed in terms of acceleration, of $+14.4~$nm.s$^{-2}$.d$^{-1}$ during gravity trend was -10.2~nm.s$^{-2}$.d$^{-1}$ and then remained stable. The initial trend due to TPLS is not enough to explain the gravity trend. Other parameters fluctuate in the CAG measurements. They are mostly due to the initial cold atom source position that can fluctuate with optical power and polarisation of other optical beams used to cool down the atoms~\cite{Farah2014b}. When released, the atomic cloud expands and the atoms sample inhomogeneous laser wavefronts distortions across the laser beam, which leads to the so-called wavefront aberration bias~\cite{LouchetChauvet2011}, \cite{Karcher2018} and so to measurement instabilities. While variations of the TPLS were the dominant source of instability for the measurements up to 2014, the implementation of laser intensity locks since then allowed to improve the long term stability. For the 27-days long measurement session presented above, this effect kept a constant value of ($39.7~\pm~0.4$)~nm.s$^{-2}$, so that we attribute the dispersion of the CFs to variations of the wavefront aberration bias.

\section{Repeated calibrations over 7 years}

\begin{figure}[h!]
\includegraphics[width=0.7\textwidth]{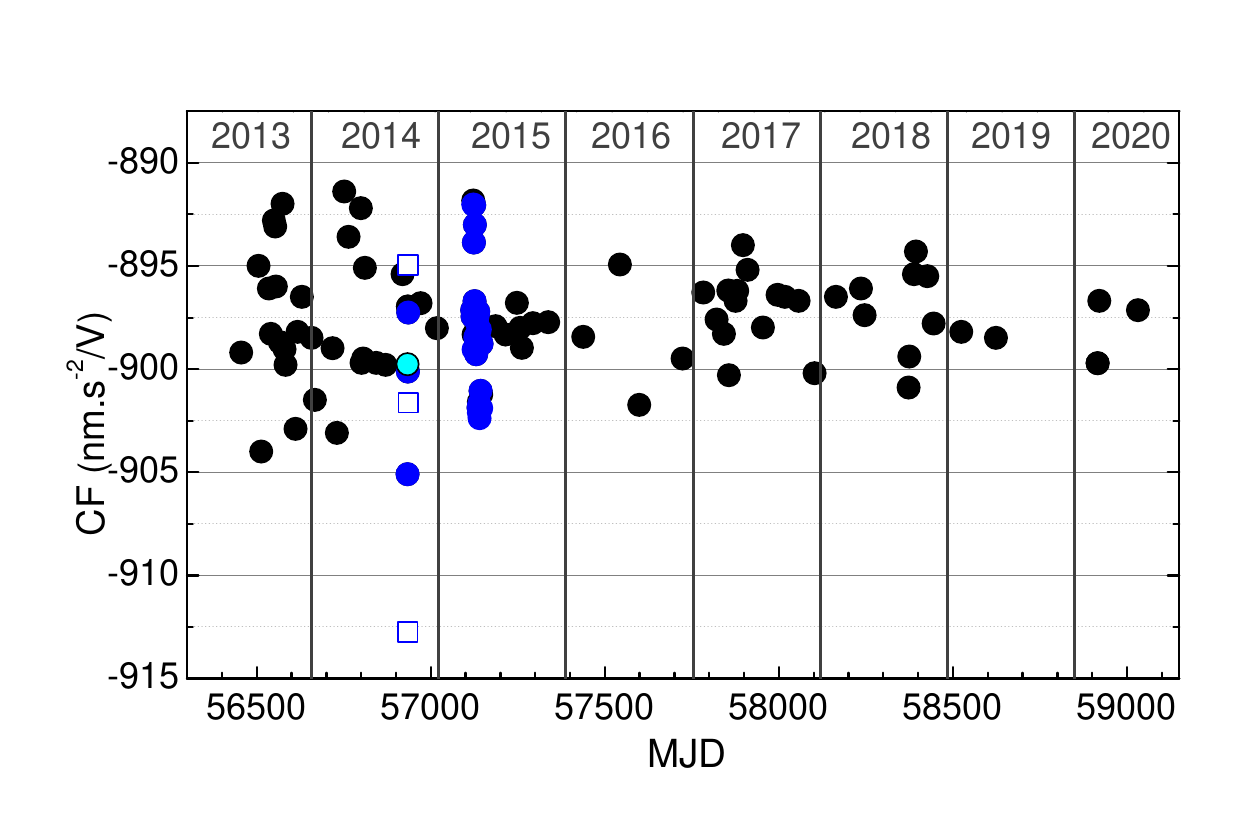}
\caption{Calibration factors of the iGrav005 obtained with the CAG since 2013 (full dots). The blue dots display the 1-day calibration CFs presented in figures~\ref{FigCFCAG1d} and \ref{Fig-CF-CAG-FG5-1d}. The opened squares display the CF obtained with the FG5X-220 in figure~\ref{Fig-CF-CAG-FG5-1d}.}
\label{FigCFlgterm}
\end{figure}

To conclude with the iGrav005 CFs, figure~\ref{FigCFlgterm} displays as black dots the results of repeated calibration campaigns realised with the CAG since 2013. The dispersion of the CFs is there comparable to the one of the CFs obtained in this paper for one-day calibrations (displayed on the same figure as blue dots), and comparable to the 15~nm.s$^{-2}$/V peak-to-peak fluctuations obtained again with FG5 in other works (over 5 months in \cite{Almavict1998} and over 10 years in \cite{Rosat2009}).

Given that we do not expect that the mean of many such determinations would be biased, we take the average of all the measurements as our best estimate of our CF with a very conservative uncertainty given by the standard deviation of the distribution. It amounts to (-897.6~$\pm$~2.7)~nm.s$^{-2}$/V.

\section{Conclusion}

We have performed the calibration of the relative SG iGrav005, using a 27-days long common view measurement with the SYRTE atomic absolute gravimeter CAG. This allowed to evaluate the long term stability of the residuals obtained by taking the difference between gravity data of the CAG and the calibrated SG. The Allan standard deviation of these residuals can reach 0.5~nm.s$^{-2}$ after averaging over close to two days for a selected quiet period, but tend to flicker at a level of about 2~-~3~nm.s$^{-2}$ when averaging over the whole period. We attribute this behavior to instabilities of CAG systematic effects rather than instabilities of the SG. By carrying out a detailed statistical analysis and a comparison with simulated data, we show how the CAG instability imposes a limit on the relative uncertainty of the determination of the SG calibration factor, of about 3\textperthousand~(1$\sigma$). This could be improved well below the \textperthousand~level with an improved long term stability of the CAG. In particular we found relative errors for the fits of the CF below the \textperthousand~level, for peak-to-peak gravity variations larger than 1~500~nm.s$^{-2}$ in a day of measurement. As an example, we demonstrated 0.5\textperthousand~after less than 2 days of measurement with a selected quiet set of data. A comparison with the calibrations realised with a corner cube FG5X gravimeter has also been performed, which shows the better performance of the CAG. Moreover, the iGrav calibration factors determined by these two types of sensors, are found in reasonable agreement but do exhibit fluctuations arising from bias instabilities.

\begin{ack}
This research is carried on within the kNOW and ITOC projects, which acknowledges the financial support of the EMRP. The EMRP was jointly funded by the European Metrology Research Programme (EMRP) participating countries within the European Association of National Metrology Institutes (EURAMET) and the European Union. B.C. thanks the Labex First-TF for financial support. This work has been supported by the Paris \^Ile-de-France R\'egion in the framework of DIM SIRTEQ.
\end{ack}

\end{document}